\documentclass[pra,twocolumn,showpacs]{revtex4}

\usepackage{graphicx}% Include figure files
\usepackage{bm,color}% bold math

\usepackage{graphicx}
\usepackage{amsmath}
\usepackage{bm,color}
\usepackage{multirow}

\newcommand{\be}{\begin{eqnarray}}
\newcommand{\ee}{\end{eqnarray}}

\usepackage{ulem}
\usepackage{xcolor}

\renewcommand{\theequation}{\arabic{equation}}

\begin{document}

\title{Interaction induced doublons and embedded topological subspace in a complete flat-band system}
\date{\today}

\author{Yoshihito Kuno}
\author{Tomonari Mizoguchi}
\author{Yasuhiro Hatsugai}

\affiliation{Department of Physics, University of Tsukuba, Tsukuba, Ibaraki 305-8571, Japan}

\begin{abstract}
In this work, we investigate effects of weak interactions on a bosonic complete flat-band system. 
By employing a band projection method, the flat-band Hamiltonian with weak interactions is mapped to an effective Hamiltonian. The effective Hamiltonian indicates that doublons behave as well-defined quasi-particles, which acquire itinerancy through the hopping induced by interactions. 
When we focus on a two-particle system, from the effective Hamiltonian, an effective subspace spanned only by doublon bases emerges. 
The effective subspace induces spreading of a single doublon and we find an interesting property: The dynamics of a single doublon keeps short-range density-density correlation in sharp contrast to a conventional two-particle spreading. 
Furthermore, when introducing a modulated weak interaction, 
we find an interaction induced topological subspace embedded in the full Hilbert space. 
We elucidate the embedded topological subspace by observing the dynamics of a single doublon, 
and show that the embedded topological subspace possesses a bulk topological invariant. 
We further expect that for the system with open boundary the embedded topological subspace has an interaction induced topological edge mode described by the doublon. The bulk--edge--correspondence holds even for the embedded topological subspace. 
%and a bulk topological index can be verified.
\end{abstract}

%\pacs{67.85.Hj, 75.10.-b, 03.75.Nt}

\maketitle
%%%%%%%%%%%%%%%%%%%%%%%%%%%%%%
\section{Introduction}
%\cintm{--Green: Corrections by TM}

Flat-band systems so far have been receiving a lot of attentions in condensed matter community. 
Physics of the flat-band is rich. Interesting subjects exist such as flat-band ferromagnetism \cite{Mielke1,Tasaki,Mielke2}, flat-band superconductors \cite{Rizzi,Tovmasyan2,Doucot,Peotta,Aoki}, and topological flat bands \cite{Katsura,Regnault,Bergholtz,Parameswaran,Neupert,Mizoguchi}, etc. 
Furthermore, localization properties of flat-band systems are exotic. Flat-band systems possesses an interesting single particle mode, ``compact localized state" \cite{Leykam}, which has a spatial spread of a few sites. 
Then, the flat-band system exhibits an exotic dynamical aspect \cite{Weimann,Mukherjee0}: A specific initial state of one particle tends to be localized dynamically. 
Furthermore, for complete flat-band systems where all bands are flat, a complete dynamical localization called Aharanov-Bohm (AB) caging is known to exist. 
Historically, a complete flat-band model was first proposed in Ref.~\onlinecite{Vidal0} and the AB caging has been expected there. 
The AB caging is also a key factor of a disorder free localization, which is a hot topic \cite{Leykam,Flach}. 
On the experimental side, two decades ago, the AB caging was first realized in GaAs/GaAlAs system \cite{Naud}. 
Recently, photonic experiments also realized a complete flat-band system \cite{Mukherjee,Kremer} and succeeded in the observation of the AB caging. 
All of these are noninteracting systems.

Here, it is very natural to ask whether or not the AB cages survive for interacting systems 
or what kind of changes the interaction gives to the AB caging. 
So far, two decades ago, the first research was done for the above asking by Vidal, {\it et.al.} \cite{Vidal1}: 
They found that, once a weak Coulomb repulsive force is switched on, 
ballistic spreading of two-particles has been expected under a flat-band diamond chain. 
In other words, interactions induce delocalization and spreading. 
This seems to be somewhat counter-intuitive. 

Even in recent theoretical studies, the fate of the AB caging under interactions has been investigated in \cite{Liberto,Zurita,Danieli1,Danieli2}, 
and the effects of interaction have been also investigated from the viewpoint of the disorder--free many--body--localization \cite{Kuno,Roy,Danieli3,Orito}. However, we have not fully elucidated the effects of the interaction for complete flat-band systems.

In this work, we investigate the fundamental effects of the interaction for a complete flat-band system by considering an interacting two-particle system on a simple one--dimensional flat-band lattice, namely flat-band Creutz ladder \cite{Creutz0, Creutz}. 
The Creutz ladder has been recently implemented in coldatoms \cite{Kang}, and also can be implemented for photonic crystals with some weak nonlinearities of the medium like Kerr nonlinearity. 
In our theoretical analysis, we make use of a band projection method \cite{Huber,Takayoshi,Bermudez1}. The Creutz ladder with weak interactions can be mapped into an effective Hamiltonian, which indicates the presence of a paired particle, namely, doublon. 
We focus on two-particle systems, and investigate characteristic properties of the single doublon, especially for its dynamical aspects and topological properties. 
Beyond the previous study \cite{Vidal1}, we will explicitly show the presence of the subspace constituted only by doublon bases 
and clearly show the interaction-induced itinerancy of single doublon in details. We find an interesting property on the complete flat-band with weak interactions, 
that is, the doublon spreading keeps a short-range density-density correlation which does not appear in a conventional two-(free)particle spreading.

Much recently, Pelegr\'i {\it et al.} \cite{Pelegri} studied a two-particle system in a diamond chain with complete flat-bands, which can be feasible in coldatoms \cite{Pelegri2}. There, they showed from the numerical diagonalization for the open system that a suitable interaction tuning induces doublon edge modes. 
In this regard, we also introduce a modulated interaction for our Creutz ladder, study the dynamical aspect, and in particular focus on the study of the bulk topology in the two-particle doublon subspace, which is a different perspective from the reference \cite{Pelegri}. 
Detection of a band topology through single particle dynamics in topological systems attracts great interest in both theory \cite{MCD3,Zhou,Haller,MKH} and experiment \cite{MCD2,MCD1}. Such a detection can be also applicable to our target system as shown in this work. 
We show that for the two-particle system a modulated interaction induces a {\it topological subspace of the doublon embedded into the full Hilbert space}, and that the topology of that subspace is also clearly visible from spreading of a single doublon, and the topological invariant can be extracted from the dynamics. 
We numerically demonstrate the dynamics of a single doublon, where the dynamics is governed only by the topological subspace, of which the Hamiltonian corresponds to the low energy effective Hamiltonian obtained by the band projection method.
Then we clearly show the presence of the bulk--edge--correspondence (BEC)~\cite{Hatsugai} for the topological subspace embedded in the full Hilbert space. 
To our knowledge, the BEC for the subspace is also a new finding. 

This paper is organized as follows.
In Sec.~II, we introduce the Creutz-ladder model with weak on-site interactions. 
Section III shows the band projection method and the effective Hamiltonian.
The properties of energy spectrum and eigenstates for two-particle system are discussed in Sec.~IV. 
Base on the properties, in Sec.~V, we clarify the dynamics of the two-particle system induced by weak uniform on-site interactions. 
In Sec.~VI, we discuss the effects of a modulated interaction. The properties of energy spectrum and eigenstates for two-particle system are discussed in detail. In particular, we show that a part of the subspace is topological. 
From the single doublon dynamics the bulk topological invariant is extracted. And we clarify the presence of the BEC. Section VII is devoted to the conclusion.

\section{Creutz ladder}
\begin{figure}[t]
\begin{center} 
\includegraphics[width=8cm]{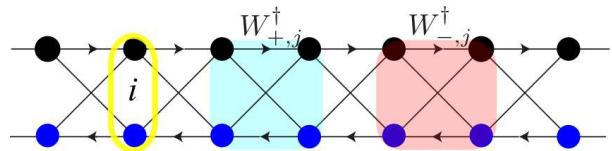} 
\end{center} 
\caption{Creutz ladder model: the yellow circle represents a unit cell, the blue and magenta shade objects are compact localized states, $W_{+,j}$ and $W_{-,j}$.}
\label{Fig1}
\end{figure}

In this work, we consider the Creutz ladder (its schematic picture is shown in Fig.~\ref{Fig1}). 
The model was originally introduced by Creutz \cite{Creutz0,Creutz}, 
whose Hamiltonian is given by 
\begin{eqnarray}
H_{\rm C}=&&\sum_{j}\biggl[-it_1(a^{\dagger}_{j+1}a_{j}-b^{\dagger}_{j+1}b_{j})\nonumber\\
&&-t_{0}(a^{\dagger}_{j+1}b_{j}+b^{\dagger}_{j+1}a_{j})\biggr],
\label{Creutz}
\end{eqnarray}
where $a^{(\dagger)}_{j}$ and $b^{(\dagger)}_{j}$ are the boson annihilation (creation) operators on the upper and lower chains, respectively. 
$j$ refers to a unit cell. $t_1$ and $t_{0}$ are the intra-chain and inter-chain hopping amplitudes, respectively. 
When $t_{0}=t_{1}$ two complete flat-bands with $E=\pm 2t_{1}$ appear due to the interference of the hoppings. 
In what follows, we set the flat-band condition, $t_{0}=t_{1}$. 
Under the flat-band condition, compact localized eigenstates of the upper and lower flat-bands are given by \cite{Takayoshi,Bermudez1}
\begin{eqnarray}
W^{\dagger}_{-, j}= \frac{1}{2}\biggl[ia^{\dagger}_{j+1}+b^{\dagger}_{j+1}
+a^\dagger_{j}+ib^{\dagger}_{j}\biggr],\\
W^{\dagger}_{+, j}= \frac{1}{2}\biggl[ia^{\dagger}_{j+1}+b^{\dagger}_{j+1}
-a^\dagger_{j}-ib^{\dagger}_{j}\biggr].
\label{CLS}
\end{eqnarray}
The $W_{\pm,j}$ operator can be regarded as bosonic particles, which satisfy
\begin{eqnarray}
[W_{\alpha, j},W^{\dagger}_{\beta,j'}]&=&\delta_{\alpha\beta}\delta_{jj'},\:\:\:
[W^{(\dagger)}_{\alpha, j},W^{(\dagger)}_{\beta,j'}]=0,\nonumber
\label{W_commutation}
\end{eqnarray}
where $[\cdot]$ implies a (bosonic) commutation relation. Accordingly, we call the particle represented by $W_{\pm,j}$ operator a ``W-particle''.
%%%%%%%%%%%%%%%%%%%%%%%%%%%%%%%%%%%%
%%%%%%%%%%%%%%%%%%%%%%%%%%%%%%%%%%%%

When one substitutes Eqs. (2) and (3) into $H_{\rm C}$, the Creutz ladder is written by
\begin{eqnarray}
H_{\rm C}=&&\sum^{L-1}_{j=0}\biggl[-2t_1 W^{\dagger}_{-,j}W_{-,j}+2t_1 W^{\dagger}_{+,j}W_{+,j}\biggr].
\label{Creutz2}
\end{eqnarray}
Here, the W-particle picture includes no hoppings and only on-site potential terms. The number operators of the W-particle $W^{\dagger}_{\pm,j}W_{\pm,j}$ is a conserved quantity. The presence of the conserved quantities leads to a disorder free localization \cite{Danieli1,Danieli2,Zurita,Kuno,Roy,Danieli3,Orito}. 
As discussed in \cite{Bermudez1}, the model also has topological properties. For a finite system with open boundary, the edge mode $\gamma_{\ell}$ ($\ell=L(R)$) appears. Then, the Creutz ladder with open boundary is written by 
\begin{eqnarray}
H_{\rm C}=&&\sum^{L-2}_{j=0}\biggl[-2t_1 W^{\dagger}_{-,j}W_{-,j}+2t_1 W^{\dagger}_{+,j}W_{+,j}\biggr]\nonumber\\
&&+\epsilon_{edge}(\gamma^{\dagger}_{L}\gamma_{L}+\gamma^{\dagger}_{R}\gamma_{R}),
\label{Creutz3}
\end{eqnarray}
where $\gamma^{(\dagger)}_{L(R)}$ is a annihilation (creation) operator of the left (right) edge mode, and 
$\epsilon_{edge}$ is an energy for the edge modes. For noninteractiong case, $\epsilon_{edge}=0$.
It should be noted that in $W$-particle sector, one lattice site is vanished due to the presence of the edge mode $\gamma_{R(L)}$. In what follows, we set $t_1=1$.

\section{Perturbation: Weak on--site interaction}
Let us introduce a weak interaction in the flat--band Creutz ladder. 
We consider the following weak on-site interaction:
\begin{eqnarray}
\delta V=\sum^{L-1}_{j=0}\biggl[
U^{a}_j a^\dagger_{j}a^\dagger_{j}a_{j}a_{j}
+U^{b}_j b^\dagger_{j}b^\dagger_{j}b_{j}b_{j}\biggl].
\label{delV}
\end{eqnarray}
In this work we assume that the interaction strength is much smaller than the band gap, $U^{a}_j,  U^{b}_j\ll|4t_1|(=4)$.

Let us express the interaction term with W-particles by using
\begin{eqnarray}
a_j&=& \frac{1}{2}\biggl[W_{-,j}-W_{+,j}
+iW_{-,j-1}+iW_{+,j-1}\biggr],\\
b_j&=& \frac{1}{2}\biggl[W_{-,j-1}+W_{+,j-1}
+iW_{-,j}-iW_{+,j}\biggr].
\label{CLS_inverse}
\end{eqnarray}
Here, we employ the band projection method \cite{Huber,Takayoshi,Bermudez1}. 
Since the interaction is weak, we assume that the $W$-particle picture is valid and that when the interaction term is expanded by W-particle operators one can ignore terms containing $W_+$ or $W_+^\dagger$. 
Under this situation, we can construct a lower flat-band projected effective Hamiltonian under periodic boundary condition, which is given as 
\begin{eqnarray}
H^{p}_{\rm e}&=&\sum^{L-1}_{j=0}\frac{U_j}{8}
\biggl[\biggl(-B^{\dagger}_{j}B_{j-1}+\mbox{h.c.}+B^{\dagger}_{j}B_{j}+B^{\dagger}_{j-1}B_{j-1}\biggr)\nonumber\\
&&+4W^{\dagger}_{-,j}W_{-,j}W^{\dagger}_{-,j-1}W_{-,j-1} \biggr]+ E_{\rm self}.
\label{Heff2}
\end{eqnarray}
Here $B^{\dagger}_{j}\equiv W^{\dagger}_{j}W^{\dagger}_{j}$, describing a doublon, $E_{\rm self}$ is a self-energy coming from the on-site interactions, and the doublon hopping terms emerge. Note that the higher flat-band projected effective Hamiltonian can also be constructed as well. 
In this work, based on the effective Hamiltonian $H^{p}_{\rm e}$, we will discuss the physics of two-particle. 

\section{Two-particle spectrum for uniform interaction}
Let us consider a uniform interaction $U^{a}_j=U^{b}_j=U$ in a two-particle system. 
This situation elucidates the essence of the interaction effects in the complete flat-band system. 
To clarify the properties of the system we calculate all energy spectrum and eigenstates for a finite system with periodic boundary condition. 

\begin{figure}[t]
\centering
%\begin{center}
%\centering  
\includegraphics[width=8cm]{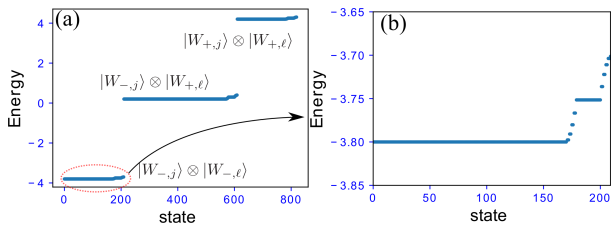}
%\end{center} 
% \hspace{10truemm}
\caption{(a) The entire energy spectrum for two-particle system. We see three bands. Here we set $L=20$, $U=0.1$. 
(b) The closeup of the lowest band spectrum. 
Between 170-th and 209-th states, a mini dispersive band structure appears, where some extended doublon eigenstates $|\phi^{B}_{\ell}\rangle$ are lurking.}
\label{Fig2}
\end{figure}

For noninteracting situation, the two-particle eigenstates are constituted just by occupying two of $W$-particle states. This picture mostly survives when 
the weak interaction is switched on. 
Figure~\ref{Fig2} (a) is the full spectrum of the two-particle system with finite $U$ \cite{Quspin}.
Here, there are three bands: approximately,
the lower band is constructed by states where two of $W^{\dagger}_{-,j}$ are occupied, 
the middle band by one $W^{\dagger}_{-,j}$ and one $W^{\dagger}_{+,j}$ are occupied,
and 
the higher band is by two of $W^{\dagger}_{+,j}$ are occupied.

%[* Comment: Product states of $W^\dagger|0\rangle$ may %be confused with $W^\dagger|0\rangle \otimes %W^\dagger|0\rangle$ ($ \neq W^\dagger W^\dagger %|0\rangle$?), thus to be avoided..]
%}

In particular, we show the closeup of the lowest band spectrum in Fig.~\ref{Fig2} (b). 
For the lowest band the effective Hamiltonian $H^{p}_{\rm e}$ is valid. 
From the form of $H^{p}_{\rm e}$ we can further separate the lowest band space by using the following three kinds of bases: 
\begin{eqnarray}
|W^{ap}_{k,\ell}\rangle &=& W^{\dagger}_{-,k} W^{\dagger}_{-,\ell}|0\rangle, \: |k-\ell|\geq 2,\\
|W^{NN}_{j}\rangle &=& W^{\dagger}_{-,j} W^{\dagger}_{-,j+1}|0\rangle,\\
|B_{j}\rangle &=&W^{\dagger}_{-,j}W^{\dagger}_{-,j}|0\rangle,
\label{two_base}
\end{eqnarray}
where $|W^{ap}_{k,\ell}\rangle$ is a two particle state that two W-particles do not overlap, 
$|W^{NN}_{j}\rangle$ is a two particle state that the two W-particles share two sites, and $|B_{j}\rangle$ is a single doublon state of W-particles; $|0\rangle$ is a vacuum. 
One can easily confirm that $|W^{ap}_{k,\ell}\rangle$ and $|W^{NN}_{j}\rangle$ become eigenstates for $H^{p}_{\rm e}$; 
\begin{eqnarray}
H^{p}_{\rm e}|W^{ap}_{k,\ell}\rangle &=& E_{\rm self}|W^{ap}_{k,\ell}\rangle, \\ 
H^{p}_{\rm e}|W^{NN}_{j}\rangle &=& \biggl(\frac{U}{2}+E_{\rm self}\biggr)|W^{NN}_{j}\rangle.
\label{two_base_eigenstates}
\end{eqnarray}
On the other hand, the doublon base $|B_{j}\rangle$ is not an eigenstate for $H^{p}_{\rm e}$ due to the presence of the doublon hopping term. In the lowest band, 
%\csout{since $|W^{ap}_{k,\ell}\rangle$ and %$|W^{NN}_{j}\rangle$ are already eigenstates for $H^{p}_{\rm e}$,} 
the doublon hopping term of $H^{p}_{\rm e}$ acts as an off-diagonal operator for $|B_{j}\rangle$. Accordingly, $H^{p}_{\rm e}$ leads to a set of the eigenstates, 
which is generally given by the linear combination of $|B_{j}\rangle$ basis: 
\begin{eqnarray}
&&|\phi^{B}_{\ell}\rangle=\sum_{j}c^{\ell}_{j}\biggl(\frac{1}{\sqrt{2}}|B_{j}\rangle\biggl),\label{phiB}\\
&&\sum_{j}|c^{\ell}_{j}|^2=1.
\end{eqnarray}
The eigenstate $|\phi^{B}_{\ell}\rangle$ are regarded as an extended state of the doublon if $c^{\ell}_{j}$ has a broad distributions with a finite value. 

Actually, such eigenstates $|\phi^{B}_{\ell}\rangle$ can be detected even if they are buried in the full set of the eigenstates. Let us introduce the following doublon subspace projector:
\begin{eqnarray}
P_B=\frac{1}{2}\sum^{L-1}_{j=0}|B_{j}\rangle \langle B_j|.
\label{PB}
\end{eqnarray}
Note that the eigenstate $|\phi^{B}_{\ell}\rangle$ are normalized such that $\langle \phi^{B}_{\ell}|P_B|\phi^{B}_{\ell}\rangle=1$. 
This means that the eigenstate $|\phi^{B}_{\ell}\rangle$ is spanned only by $|B_{j}\rangle$ basis. 
Numerically, we calculated $P_B$ for all of the lowest band eigenstates. 
The result is shown in Appendix. We can numerically find $L$ eigenstates spanned only by $|B_{j}\rangle$ basis. The set of the eigenstates $|\phi^{B}_{\ell}\rangle$ is embedded in the full Hilbert space.
In what follows we call the set of the eigenstates ``doublon subspace''. The dimension of the doublon subspace is $L$.

%%%%%%%%%%%%%%%%%%%%%%%%%%%%%%%%%%%%
%Fig
%\widetext
%%%%%%%%%%%%%%%%%%%%%%%%%%%%%%%%%%%%

\begin{figure}[t]
\centering
%\begin{center}
%\centering  
\includegraphics[width=7.5cm]{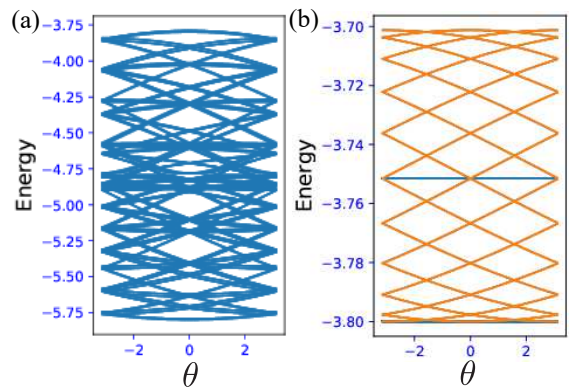}
%\end{center} 
% \hspace{10truemm}
\caption{Flux insertion response: (a) nonflat-band case, $t_0=1.5$, $U=0.1$, (b) Flat-band case, $t_0=1$, $U=0.1$. 
The orange lines represent the eigenstates constituted only by the doublon bases $|B_{j}\rangle$, that is $|\phi^{B}_{\ell}\rangle$. The number of the orange spectrum lines is $L$.}
\label{Fig3}
\end{figure}

We further characterize the doublon subspace. Since the effective Hamiltonian $H^{p}_{\rm e}$ possesses the doublon hopping term, we expect that the hopping term induces extended eigenstates of the doublon. 
Let us verify whether or not such an extended state exists as the subspace.
To this end, we observe the response to a flux insertion \cite{Gladchenko,Mondaini,Tovmasyan1}, which can be introduced by a phase twist for the hopping amplitude: 
$t_0\to t_0\exp(i\theta/L)$, $t_1\to t_1\exp(i\theta/L)$ \cite{Barbarino}. 
When one varies $\theta$ from $-\pi$, the single magnetic flux is injected into the periodic system. 
As for the general response of the flux insertion, the followings are known \cite{Gladchenko,Mondaini,Tovmasyan1}: 
While an eigenstate composed of single particles exhibits $2\pi$--periodicity, an eigenstate constituted by a paired object bases exhibits $\pi$--periodicity since the charge of the paired object is twice as large as that of a single particle. We observe the flux response for the lowest band spectrum of the two-particle system. 
The numerical results are shown in Fig.\ref{Fig3} (a) and (b). Here, for comparison, we also investigate the case of breaking the flat-band condition (i.e., $t_0 \neq t_1$). 
For nonflat-band and noninteracting case, all spectrum in the lowest band exhibit $2\pi$--periodicity. On the other hand, for flat-band and weak interaction case (Fig.\ref{Fig3} (b)), interestingly enough some spectrum exhibit $\pi$-periodicity while the others do not response to the flux insertion. Actually, the orange spectrum lines in Fig.\ref{Fig3} (b)  correspond to eigenstates constituted by the doublon base, that is, they belong to the doublon subspace. On the other hand, the blue lines in Fig.\ref{Fig3} (b) are much degenerate, and the corresponding eigenstates are somewhat localized. Such states do not respond to the flux insertion \cite{Filippone}. These states are constituted by $|W^{ap}_{k,\ell}\rangle$ and $|W^{NN}_{j}\rangle$ bases. 
From the response to the flux insertion, we confirm that the extended eigenstates constituted by the doublon base exist within the full Hilbert space. So far such a response has been reported as for the many-body groundstate \cite{Mondaini,Tovmasyan1}, 
but in our data, we observed that excited states also exhibit such $\pi$-periodicity.   

%%%%%%%%%%%%%%%%%%%%%%%%%%%%%%%%%%%%
%Fig
%\widetext
%%%%%%%%%%%%%%%%%%%%%%%%%%%%%%%%%%%%

\section{Unitary dynamics for uniform interaction}
So far, we have clarified the spectrum and eigenstates for the two-particle system with the uniform interaction.  
In this section, we show that the two-particle system governed by $H^{p}_{\rm e}$ exhibits interesting dynamics. 
From $H^{p}_{\rm e}$, the doublon hopping terms make the doublon state extended. 
Then, if one puts a single doublon at $j$, what dynamics does the initial single doublon show? 
Without interaction, 
the doublon initial state is just two noninteracting $W$-particles.
On the flat-band system, each particle does not spread out, that is, exhibit just the AB caging. 
On the other hand, for finite $U$, where the flat-band is partially distorted, 
some eigenstates turn into extended states while the rest of the eigenstates are localized states. 
From such extended states, if one puts a single doublon as initial state, the dynamics is expected to exhibit spreading. 
Interestingly enough, the dynamics keeps a correlation. 
In what follows, we discuss such a single doublon dynamics and show the explicit analytical description for it, 
and numerically demonstrate the dynamics of the single doublon.

%%%%%%%%%%%%%%%%%%%%%%%%%%%%%%%%%%%%
%Fig
%\widetext
\begin{figure*}[t]
\centering
%\begin{center}
%\centering  
\includegraphics[width=18cm]{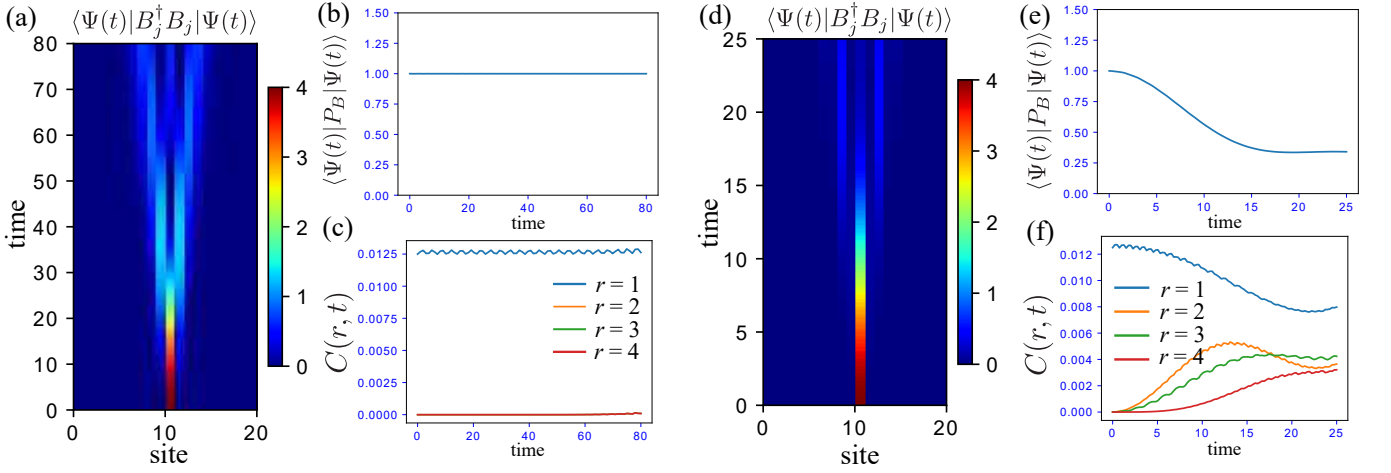}
%\end{center} 
% \hspace{10truemm}
\caption{(a) Time evolution of doublon density for flat-band case ($U=0.1$, $t_0=1$). The initial state is $B_{j_0}|0\rangle$.
(b) The dynamical behavior of the eigenvalue of the doublon projector, $\langle \Psi(t)| P_{B}|\Psi(t)\rangle$ for flat-band case.
(c) The dynamical behavior of the density-density correlation $C(r,t)$ for flat-band case.
(d) Time evolution of doublon density for nonflat-band case ($U=0.1$, $t_0=1.08$). The initial state is $B_{j_0}|0\rangle$. 
(e) The dynamical behavior of the eigenvalue of the doublon projector, $\langle \Psi(t)| P_{B}|\Psi(t)\rangle$ for nonflat-band case.
(f) The dynamical behavior of the density-density correlation $C(r,t)$ for nonflat-band case.}
\label{Fig4}
\end{figure*}
%%%%%%%%%%%%%%%%%%%%%%%%%%%%%%%%%%%%

We investigate the dynamics of the initial state $|B_{j_0}\rangle$, where $j_{0}=(L-2)/2$. 
The dynamics is given by a unitary dynamics. 
Here, we recognize that there are the following relations for the two W-particle bases:
\begin{eqnarray}
\langle B_{j}|B_{j'}\rangle=2\delta_{jj'},\:
\langle B_{j}|W^{NN}_{j'}\rangle=0,\:
\langle B_{j}|W^{ap}_{k,\ell}\rangle=0.
\label{relations}
\end{eqnarray}
Furthermore, since $W_+$ and $W_-$ are orthogonal to each other at single-particle level, all eigenstates in the middle and higher bands described by $|\phi^{\rm{mh}}_{\ell}\rangle$ is orthogonal to $|B_{j}\rangle$, $\langle \phi^{\rm{mh}}_{\ell}|B_j\rangle=0$.
Then, the unitary dynamics can be given by the following form (we set $\hbar=1$)
\begin{eqnarray}
&&U(t)=U^{L}(t)+U^{\rm{mh}}(t),
\label{unitary_daynamics}
\end{eqnarray}
where 
\begin{eqnarray}
U^{L}(t)&=&\sum_{\ell}e^{-i\epsilon^{B}_{\ell}t}|\phi^{B}_{\ell}\rangle \langle\phi^{B}_{\ell}|\nonumber\\
&+&\sum_{\ell}e^{-i\epsilon^{NN}_{\ell}t}|W^{NN}_{\ell}\rangle \langle W^{NN}_{\ell}|\nonumber\\
&+&\sum_{\ell}e^{-i\epsilon^{ap}_{\ell}t}|W^{ap}_{\ell}\rangle \langle W^{ap}_{\ell}|,\\
U^{\rm{mh}}(t)&=&\sum_{\ell}e^{-i\epsilon^{\rm{mh}}_{\ell}t}|\phi^{\rm{mh}}_{\ell}\rangle \langle \phi^{\rm{mh}}_{\ell}|.
%|\phi^{B}_{\ell}\rangle&=&\sum^{L-1}_{j=0}c^{\ell}_{j}\biggl(\frac{1}{\sqrt{2}}|B_{j}\rangle\biggl).
\label{unitary_daynamics_v2}
\end{eqnarray}
Here, $\epsilon^{b}_{\ell}$, $\epsilon^{NN}_{j}$ and $\epsilon^{ap}_{\ell}$ are eigenvalues of the eigenstates for the subspace of the lowest band, and $\epsilon^{\rm{mh}}_{\ell}$ is an eigenvalue of the eigenstates in the middle and higher band. 

Due to the orthogonal relations in Eq.~(\ref{relations}) and $\langle \phi^{\rm{mh}}_{\ell}|B_j\rangle=0$, 
the unitary dynamics for the initial state $|B_{j_0}\rangle$ is governed only by $U^{L}$, the dynamics is given as a simple form: 
\begin{eqnarray}
|\Psi(t)\rangle &=&U(t)|B_{j_0}\rangle =U^{L}(t)|B_{j_0}\rangle \nonumber\\
&=&\sum_{\ell}e^{-i\epsilon^{b}_{\ell}t}|\phi^{B}_{\ell}\rangle \langle \phi^{B}_{\ell}|B_{j_0}\rangle,
\label{unitary_daynamics_B0}
\end{eqnarray}
that is, the dynamics of the single doublon is governed only by the doublon subspace, i.e., the set of the eigenstates $|\phi^{B}_{\ell}\rangle$. The extended eigenstate $|\phi^{B}_{\ell}\rangle$ makes the initial localized single doublon spread along the time evolution.
Accordingly, the weak interaction $U$ induces spreading phenomena on the flat-band system, which is somewhat counter-intuitive. 

We further show that the dynamics of the single doublon possesses an interesting correlation property. 
Let us introduce a density-density correlation:
\begin{eqnarray}
C(r,t)=\frac{1}{L}\sum_{i}\langle \Psi(t)|n^{\alpha}_{i}n^{\alpha}_{i+r}|\Psi(t)\rangle
\label{unitary_daynamics_B0}
\end{eqnarray}
where $n^{\alpha}_{i}=a^{\dagger}_{i}a_{i}$ ($\alpha=a$), $(b^{\dagger}_{i}b_i)$ $(\alpha=b)$, $i$ is a lattice site for each chain in the Creutz ladder. In what follows, we set $\alpha=a$. One can show that $C(r,t)$ has a specific property by the following calculation.

To begin with, we remark that the following conditions hold:
\begin{eqnarray}
\langle B_{j}|n^{a}_{i}|B_{j}\rangle=
\begin{cases}
    1 & (i\in \tilde{j}) \\
    0 & (otherwise)
  \end{cases}
  \label{BnnB}
\end{eqnarray}
where $\tilde{j}$ is the set of sites resided in the doublon at $j$, $|B_{j}\rangle$ (there are four sites in a set $\tilde{j}$).
From this relation of Eq.~(\ref{BnnB}), the following conditions are obtained:
\begin{eqnarray}
\langle B_{j_1}|n^{a}_{i_1}n^{a}_{i_2}|B_{j_2}\rangle
=
\begin{cases}
    \delta_{j_1 j_2}\times (\mbox{const.}) & (i_1, i_2\in \tilde{j_1})\\
    0 & (\mathrm{otherwise}) 
  \end{cases}.\nonumber \\
  \label{BnnBcase}
\end{eqnarray}
Let us consider $i_1+1=i_2$ ($r=1$) case. Then the condition of Eq.~(\ref{BnnBcase}) becomes
\begin{eqnarray}
\langle B_j|n^{a}_{i_1}n^{a}_{i_1+1}|B_{j'}\rangle=\frac{1}{4}\delta_{jj'}, \:\: (i_1, i_1+1\in \tilde{j}),
\label{BnnB2}
\end{eqnarray}
where $i_1, i_1+1\in\tilde{j}$.
By using the above condition of Eq.~(\ref{BnnB2}) and the eigenstates $|\phi^{B}_{\ell}\rangle$ of Eq.~(\ref{phiB}), we can express the short-range correlation, $C(r=1,t)$ as
\begin{eqnarray}
C(1,t)&=&\frac{1}{L}\sum_{i_1}\langle \Psi(t)|n^{a}_{i_1}n^{a}_{i_1+1}|\Psi(t)\rangle\nonumber\\
&=&\frac{1}{L}\sum_{i_1,\ell,\ell',j,j'}
K^{\ell' *}_{j'}(t)K^{\ell}_{j}(t)\langle B_j|n^{a}_{i_1}n^{a}_{i_1+1}|B_{j}\rangle\nonumber\\
&=&\frac{1}{4L}\sum_{j_1,\ell,\ell'}K^{\ell*}_{j_1}(t)K^{\ell'}_{j_1}(t),
\end{eqnarray}
where 
\begin{eqnarray}
K^{\ell}_{j}(t)&=&e^{-i\epsilon^{b}_{\ell}t}c^{\ell}_j c^{\ell *}_{j_0}.\label{Kellj}
\end{eqnarray}
From this form, $C(1,t)$ has possibility to take a finite value with oscillation coming from the exponential factor in Eq.~(\ref{Kellj}).
On the other hand, for $r\geq 2$, we expect that the correlation suddenly vanishes, $C(r,t)=0$, 
due to the condition of Eq.~(\ref{BnnBcase}). 
Therefore, the density-density correlation behaves as
\begin{eqnarray}
C(r,t)=
\begin{cases}
    (\mbox{finite}) & r\leq 1  \\
    0 & (r\geq 2)
  \end{cases}.
\label{Crt_pre}
\end{eqnarray}
This implies that the unitary dynamics of the single doublon possesses the short-range density-density correlation, but no long-range one during the time evolution. Note that for spreading of free two particles, $C(r,t)$ has possibility to have a finite value even for $r\geq 2$ during the time evolution.

%%%%%%%%%%%%%%%%%%%%%%%%%%%%%%%%%%%%
%Fig
%\widetext
\begin{figure}[t]
\centering
%\begin{center}
%\centering  
\includegraphics[width=8.5cm]{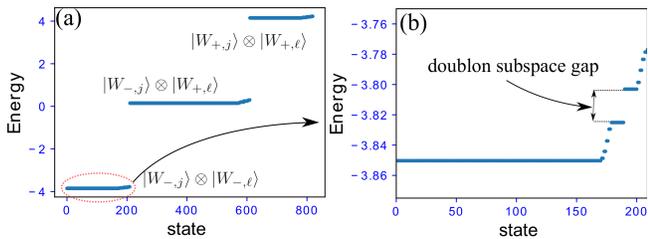}
%\end{center} 
% \hspace{10truemm}
\caption{(a) The entire energy spectrum, we see three bands. Here we set $L=20$, $U_e=0.1$ and $U_{e}=0.05$. (b) The closeup of the lowest band spectrum. The band splits to small sub-bands due to the modulated on-site interaction. The small band gap originates from the doublon subspace.}
\label{Fig6}
\end{figure}
%%%%%%%%%%%%%%%%%%%%%%%%%%%%%%%%%%%%

Let us turn to numerical demonstrations.
As characteristic dynamics, the followings are expected: 
\begin{itemize}
  \item If the doublon spreading occurs, the eigenvalue of $P_B$ keeps almost unity during time evolution. 
  \item Even under the doublon spreading, the density-density correlation function $C(r,t)$ keeps short-range correlations.
\end{itemize}

Figure \ref{Fig4} is the numerical results for the flat-band case. For the calculation of the dynamics, we set the unit of time as $\hbar/t_1=1$.
We first calculated the density of the doublon defined by $\langle\Psi(t)|B^{\dagger}_{j}B_{j}|\Psi(t)\rangle$ for each site $j$ during time evolution. The result is shown in Fig.~\ref{Fig4} (a). Here, we remark that $\langle\Psi(t)|B^{\dagger}_{j}B_{j}|\Psi(t)\rangle=4$ means the single doublon localizes at the site $j$. 
For the weak interaction the initial localized single doublon exhibits an isotropic ballistic expansion, and also the expectation value of the doublon projector $P_B$ stays unity during time evolution as shown in Fig.~\ref{Fig4} (b). This indicates that spreading of the single doublon occurs with the doublon pairing retained. As shown in Fig.~\ref{Fig4} (c) the numerical results for the density-density correlation $C(r,t)$ is consistent with the analytical prediction of Eq.~(\ref{Crt_pre}), that is, the short--range correlation $C(1,t)$ is only finite during the time evolution and the more longer correlations are zero. 

On the other hand, for nonflat-band case, the initial single doublon spreads and decays as shown in Fig.~\ref{Fig4} (d), and the expectation value of $P_B$ decreases during time evolution, as shown in Fig.~\ref{Fig4} (e). 
The initial doublon disappears with spreading, and also the density-density correlation $C(r,t)$ behaves differently from the flat-band case, the long-range correlations develop during time evolution [Fig.~\ref{Fig4} (f)]. 

\section{Emergent topological subspace}
In this section, we consider spatially--modulated on-site interaction. 
The modulation makes the band structure of the doublon subspace gapped. 
The gapped doublon can subspace exhibit non-trivial topology. 
From now on, we discuss the origin of the topological character and numerically-demonstrate the topological properties. 

%%%%%%%%%%%%%%%%%%%%%%%%%%%%%%%%%%%%
%Fig
%\widetext
\begin{figure}[t]
\centering
%\begin{center}
%\centering  
\includegraphics[width=8.5cm]{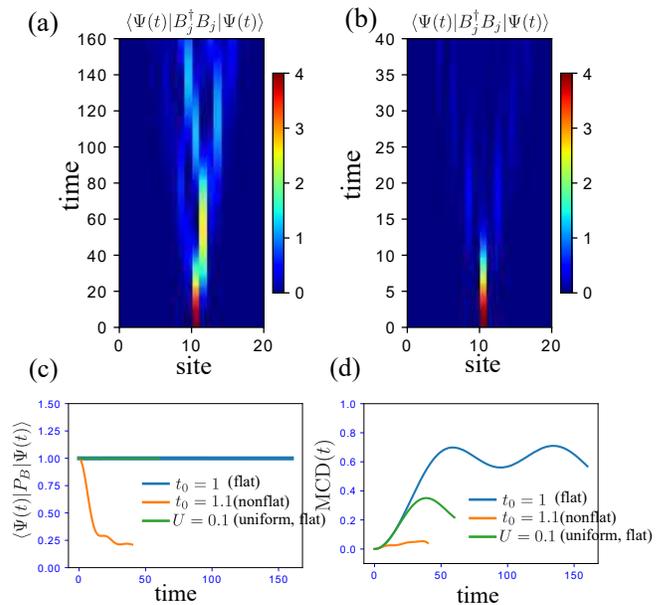}
%\end{center} 
% \hspace{10truemm}
\caption{Time evolution of doublon density for flat-band (a) and nonflat-band ($t_{0}=1.1$) (b). (c) The behavior of the doublon projector, $\langle \Psi(t)| P_{B}|\Psi(t)\rangle$. 
(d) Time evolution of the doublon MCD for various parameter cases.}
\label{Fig7}
\end{figure}
%%%%%%%%%%%%%%%%%%%%%%%%%%%%%%%%%%%%

Let us set $U^{a}_{j\in odd(even)}=U^{b}_{j\in odd(even)}=U_{o(e)}$ and consider two-particle system.  
In this case, the band projection method is valid and the lower band effective Hamiltonian only for the doublon base $|B_{j}\rangle$ is given by
\begin{eqnarray}
H^{pB}_{\rm e}=\sum^{L-1}_{j=0}\frac{U_j}{8}\biggl(-B^{\dagger}_{j}B_{j-1}+\mbox{h.c.}+B^{\dagger}_{j}B_{j}+B^{\dagger}_{j-1}B_{j-1}\biggr),\nonumber\\
\label{Heff2_B}
\end{eqnarray}
If we focus only on the doublon subspace, 
the above effective Hamiltonian can be regarded as the Su-Schrieffer-Heeger (SSH) model~\cite{Su}. 
The spectrum described by $H^{pB}_{\rm e}$ can be regarded as a single particle spectrum for doublon. 
Figure~\ref{Fig6} (a) is the full spectrum for the two-particle system. Globally, three bands appear again. 
But, if one looks at the lowest band in more detail, the lowest band exhibits a small gap as shown in Fig.~\ref{Fig6} (b). 
Here, the doublon subspace is embedded in the lowest band spectrum; its numerical verification is shown in Appendix. 
The small band gap in the lowest band comes from the gapped doublon subspace. 
The gap size is $\frac{|U_e-U_o|}{8}\times 2$ \cite{doublon_field} and also there are not any other eigenstates of the different subspace in the gap. 

The bulk topological invariant of the doublon subspace can be extracted from the dynamics of the single doublon. 
As is the case with the uniform interaction, we set a single doublon state as an initial state for dynamics.
Then, the unitary dynamics is governed only by the eigenstates of the doublon subspace described by $H^{pB}_{\rm e}$ since all other eigenstates in the Hilbert space are orthogonal to the doublon states. 
The bulk topology of the doublon subspace is expected to be characterized by the winding number. 
To extract the bulk topology for the doublon subspace, 
we introduce the doublon mean chiral displacement (MCD) \cite{MCD1,MCD2,MCD3}, which can extract the winding number  from the dynamics of a single doublon. 
The doublon MCD for the doublon subspace can be given by
\begin{eqnarray}
&&{\rm MCD}(t)=\langle \Psi(t)|\Gamma|\Psi(t)\rangle,\\
&&\Gamma=\frac{1}{4}\sum^{L/2-1}_{m=0}m\biggl[|B_{2m}\rangle \langle B_{2m}|-|B_{2m+1}\rangle \langle B_{2m+1}|\biggr],\nonumber\\
\label{Heff2_B}
\end{eqnarray}
where $\Gamma$ is a multiple operator created by the position and chiral operator for the doublon.
If the doublon subspace has a non-trivial topology, ${\rm MCD}(t)$ oscillates around $1/2$ for a long time, and 
if the long-time average is close to $1/2$, 
it is the signal that the winding number of the doublon subspace is one \cite{MCD1,MCD2,MCD3,Zhou,Haller}. 
We note that the winding number also corresponds to the Zak phase \cite{Maruyama}. For trivial topology, the ${\rm MCD}(t)$ oscillates around zero. This means that the winding number is zero.

Let us show numerical results, we investigate the unitary dynamics of the initial state $|B_{j_0}\rangle$, where $j_{0}=(L-2)/2$. We set $U_o=0.1$, $U_e=0.05$. 
%\csout{Figure~\ref{Fig7} is the results of the dynamics.} 
The density of the doublon is shown in Fig.~\ref{Fig7} (a) and (b). For the flat-band case anisotropic spreading of the doublon appears while for the nonflat-band case, the density of the doublon decays with spreading during time evolution. The expectation values of $P_B$ for various cases are shown in Fig.~\ref{Fig7} (c). For the flat-band cases, the values of $P_B$ stay unity, the doublon paring is preserved, but for the nonflat-band case, the value of $P_B$ exhibits sudden decrease, meaning that the doublon suddenly decays during time evolution. The result of ${\rm MCD}(t)$ is shown in Fig.~\ref{Fig7} (d), under the flat-band and modulated interaction case, the value of ${\rm MCD}(t)$ exhibits oscillation around $1/2$, this is the signal that the doublon subspace possesses the non-trivial topology, that is, the winding number $\gamma_{\rm w}=1$. For other cases, such a doublon MCD behavior does not appear, that is, the system is in trivial phase. From these, the doublon subspace under the modulated interactions has a non-trivial topology characterized by the winding number.

%%%%%%%%%%%%%%%%%%%%%%%%%%%%%%%%%%%%
%Fig
%\widetext
\begin{figure}[t]
\centering
%\begin{center}
%\centering  
\includegraphics[width=8.5cm]{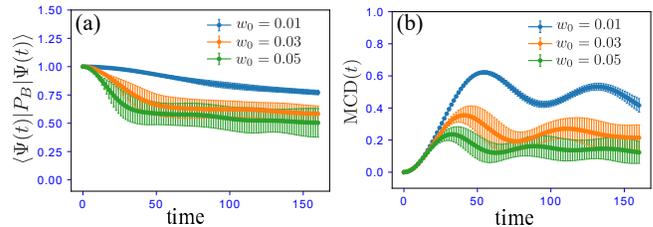}
%\end{center} 
% \hspace{10truemm}
\caption{(a) Disorder-averaged expectation value of $P_B$ during time evolution. (b) Disorder-averaged doubon ${\rm MCD}$. we set $L=20$, and used 300 quench disorder samples.}
\label{Fig8}
\end{figure}
%%%%%%%%%%%%%%%%%%%%%%%%%%%%%%%%%%%%

\subsection{Robustness for weak disorder}
Let us investigate the robustness for the topological properties for the doublon subspace. 
We introduce a uniform disorder for the interaction as $U^{a}_{j\in odd(even)}=U^{b}_{j\in odd(even)}=U_{o(e)}\to U_{o(e)}+\delta w_j$, where $\delta w_j\in [-w_0,w_0]$. 
How do the doublon projector and the doublon MCD behave? Figure~\ref{Fig8} shows the effects of (quenched) disorder for $P_B$ and ${\rm MCD}(t)$. For $P_B$, for weak disorder $w_0=0.01$, 
the value of $P_B$ is a little suppressed however, it does not decay so much during time evolution. 
On the other hand, for more large disorder $w_0=0.03$ and $0.05$ larger than the bulk gap of the doublon subspace, 
the value of $P_B$ decays to some extent, the saturation value seems to be $\sim 0.5$.
Accordingly, the robustness of $P_B$ for a disorder is qualitatively determined by the subspace bulk gap. 
This is consistent with the behavior of ${\rm MCD}$. As shown in Fig.~\ref{Fig8} (b), 
for a weak disorder, ${\rm MCD}(t)$ is not suppressed and almost oscillate around $1/2$. It implies that the doublon subspace still has non--trivial topology. On the other hand, for larger disorder $w_0=0.03$ and $0.05$, ${\rm MCD}(t)$ are highly suppressed, it implies that the doublon subspace no longer has  non-trivial topology.

\subsection{Topological doublon edge mode}
We observed that the doublon subspace has the bulk topological invariant, characterized by the dynamics of a doublon. 
According to the spirit of the BEC, the bulk topology in the periodic system induces some edge modes under open boundary condition~\cite{Hatsugai}. Here, we numerically verify the presence of edge states of the doublon and BEC for the doublon subspace.

\begin{figure}[t]
\centering
%\begin{center}
%\centering  
\includegraphics[width=8.5cm]{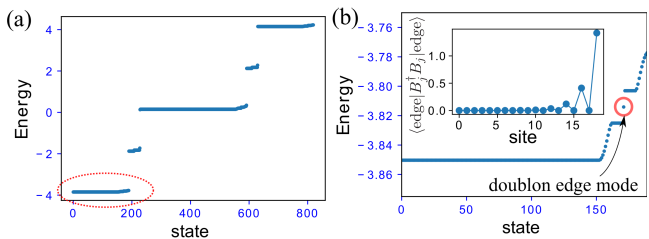}
%\end{center} 
% \hspace{10truemm}
\caption{(a) The entire energy spectrum, there are five bands. 
(b) The lowest band spectrum. The band splits to the mini sub-band due to the modulated on-site interaction. 
Within the band gap, there appears a single edge mode (the red circle) described by the doublon. The inset shows the doublon density for the edge mode. Here,we set $L=20$, $U_o=0.1$ and $U_e=0.05$.
}
\label{Fig9}
\end{figure}

When we consider the system with open boundaries, as shown in Eq.~(\ref{Creutz3}), the original Creutz ladder possesses the single--particle edge modes. 
The W-particle $W_{\pm,j}$ residing at edges turn into left and right edge modes. Let us look for an edge mode  different from the single-particle ones. 
The effective Hamiltonian by using the band projection method is given by 
\begin{eqnarray}
H^{o}_{\rm e}&=&\sum^{L-2}_{j=0}\frac{U_j}{8}
\biggl[\biggl(-B^{\dagger}_{j}B_{j-1}+\mbox{h.c.}+B^{\dagger}_{j}B_{j}+B^{\dagger}_{j-1}B_{j-1}\biggr)\nonumber\\
&&+4W^{\dagger}_{j}W_{j}W^{\dagger}_{j-1}W_{j-1}\biggr]+E_{\rm self}.
\label{Heff2open}
\end{eqnarray}
Here, if one focuses on the two-particle case and the doublon subspace, the model corresponds to the SSH model described by $B_j$ with open boundaries. If $L$ is even, the SSH chain has an {\it odd} number of the total site. This leads to a single edge mode localized at either edge. Let us numerically treat the two-particle system with open boundaries. The entire spectrum of the two-particle system is shown in Fig.~\ref{Fig9} (a). Compared to the periodic case in Fig.~\ref{Fig6} (a), the open boundary case exhibits five bands. This is due to the presence of the original (single particle) edge states $\gamma^{\dagger}_{R(L)}$ with zero energy. And for the lowest band as shown in Fig.~\ref{Fig9} (b), we find that a single edge mode resides within the small gap induced by the doublon SSH model in $H^{o}_{\rm e}$. See the inset in Fig.~\ref{Fig9} (b), 
the density distribution of doublon certainly reflects an edge state, given by $|{\rm edge}\rangle \propto [B^{\dagger}_{L-2}+(U_{e}/U_o)^{1}B^{\dagger}_{L-4}+(U_{e}/U_o)^{2}B^{\dagger}_{L-6}+\cdots]|0\rangle$, 
which can be regarded as a doublon edge mode. 
We emphasize that the doublon edge mode is different from the signle-particle edge mode since the doublon edge mode is induced by the modulated interactions and is also a two-particle object.

From these facts, we conclude that the BEC holds for the doublon subspace embedded in the full Hilbert space.

%%%%%%%%%%%%%%%%%%%%%%%%%%%%%%%%%%%%
%Fig
%\widetext
%%%%%%%%%%%%%%%%%%%%%%%%%%%%%%%%%%%%
%%%%%%%%%%%%%%%%%%%%%%%%%%%%%%%%%%%%
%Fig
%\widetext
\begin{figure*}[t]
\centering
%\begin{center}
%\centering  
\includegraphics[width=14cm]{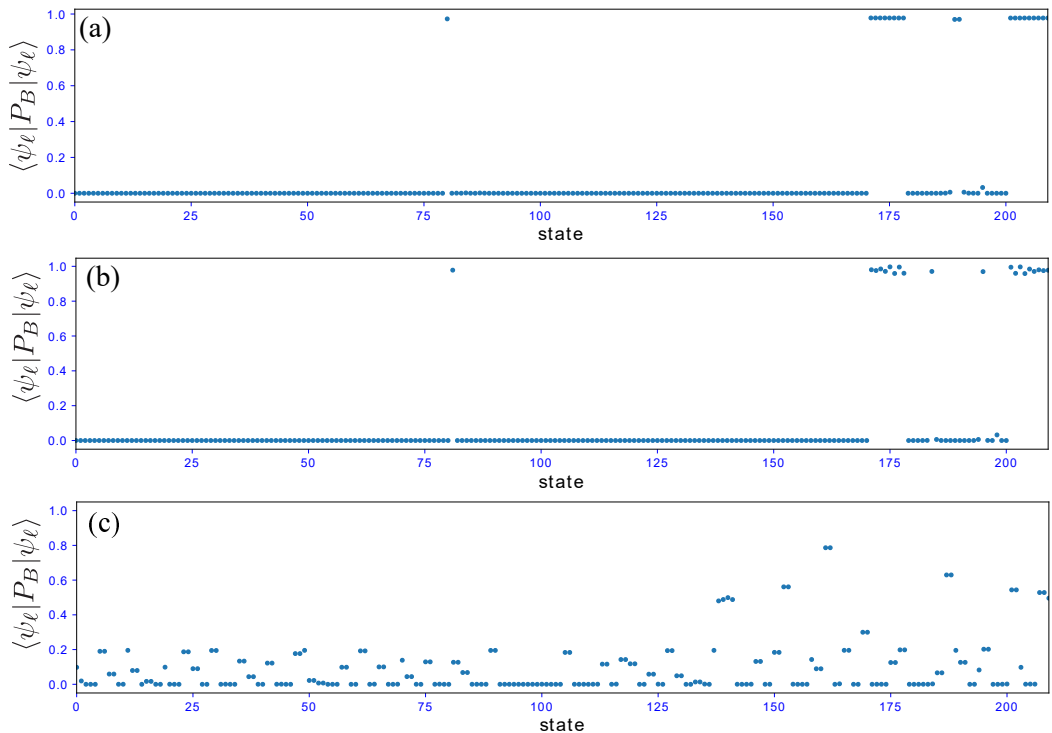}
%\end{center} 
% \hspace{10truemm}
\caption{Distribution of eigenvalues of the doublon projector $P_B$. 
(a) Flat-band and uniform interaction, $t_0=1$, $U=0.1$. 
(b) Flat-band and modulated interaction, $t_0=1$, $U_o=0.1$, $U_=0.05$.
(c) Dispersive band and uniform interaction, $t_0=1.08$, $U=0.1$.
For all cases, $L=20$ two-particle system, and we display the states in the lowest band.}
\label{FigA1}
\end{figure*}
%%%%%%%%%%%%%%%%%%%%%%%%%%%%%%%%%%%%

\section{Conclusion}
In this work, we considered two-particle physics in a complete flat-band model with weak on-site interactions.
The weak interactions induce a subspace spanned by doublon bases, which is embedded in the full Hilbert space. 
The eigenstates of the doublon subspace are extended. This fact leads to spreading for a single doublon. 
The spreading has an interesting property: 
the density-density correlation exhibits short-range correlations. 
Furthermore, we clarified that a spatially modulated weak interaction induces a topological subspace described by the doublon bases. 
Numerically, we showed that the bulk topological invariant for the doublon subspace can be extracted by observing the unitary dynamics of the single doublon and then, for open boundary case, we have found a doublon edge mode which is different from the original topological edge modes in the noninteracting Creutz ladder. 
Accordingly, we clearly showed that the topological doublon subspace embedded in the full-Hilbert space exhibits the BEC.
We hope these findings will be verified in future experiments. 
In this work, we considered repulsive interactions for the Creutz ladder, but we expect that interaction-induced itinerant doublons and their topology will appear in broader class of models, including attractively interacting systems.

\section*{Acknowledgments}
The work is supported in part by JSPS
KAKENHI Grant Numbers JP17H06138 (Y.K, Y.H.) and JP20K14371 (T.M.).
%%%%%%%%%%%%%%%%%%%%%%%%%%%%%%%%%%%%%%%%%%%%%%%

\renewcommand{\thesection}{A\arabic{section}} 
\renewcommand{\theequation}{A\arabic{equation}}
\renewcommand{\thefigure}{A\arabic{figure}}
\setcounter{equation}{0}
\setcounter{figure}{0}
%\section*{Appendix}
%\section*{Supplemental Material}
%\section*{\large{Supplemental Material}}
\section*{Appendix: Eigenvalues of doublon projector}
We plot the eigenvalues of the doublon projector $P_{B}$ for eigenstates for two-particle system. 
Figure.~\ref{FigA1} is the result of the eigenvalues of the doublon projector $P_{B}$ for all eigenstates in the lowest band.
The result for flat-band and uniform interaction indicates that $L$ eigenstates have the nonzero eigenvalue 1, the others have zero eigenvalue [Fig.~\ref{FigA1} (a)]. The former eigenstates are composed only by the doublon bases $|B_j\rangle$. For the modulated interaction case shown in Fig.~\ref{FigA1} (b), the results also exhibit $L$ eigenstates having the nonzero eigenvalue almost 1. On the other hand, for nonflat-band and uniform interaction case shown in Fig.~\ref{FigA1} (c), there is no eigenstate having the nozero eigenvalue 1.


\begin{thebibliography}{99}
\bibitem{doublon_field}
The factor 2 comes from the fact that the doublon base $|B_{j}\rangle$ is not normalized as Eq.~(\ref{relations}).
\end{thebibliography}
\end{document}